# Decomposition of Kolmogorov Complexity And Link To Geometry

Dara O Shayda

May 2012

## Abstract

A link between Kolmogorov Complexity and geometry is uncovered. A similar concept of projection and vector decomposition is described for Kolmogorov Complexity. By using a simple approximation to the Kolmogorov Complexity, coded in *Mathematica*, the derived formulas are tested and used to study the geometry of Light Cone.

**Keywords**: Kolmogorov Complexity, Light Cone, Lightcone, Geometry, High-Pass Filter, Low-Pass Filter.

## Motivation

**Max Planck**
(1858-1947):   Eine neue wissenschaftliche Wahrheit pflegt sich nicht in der Weise durchzusetzen, daß ihre Gegner überzeugt werden und sich als belehrt erklären, sondern vielmehr dadurch, daß ihre Gegner allmählich aussterben und daß die heranwachsende Generation von vornherein mit der Wahrheit vertraut geworden ist.

[A new scientific truth does not triumph by convincing its opponents and making them see the light, but rather because its opponents eventually die, and a new generation grows up that is familiar with it.]

## 1. Decomposition

**Definition 1.1**: A sequence of computable functions $\{f_i\}_{i=1}^{\infty}$ is called bounded if the $K(f_i) \leq M$ for some positive number M. M is called Program Upper Bound.

**Definition 1.2**: A sequence of bounded computable functions $\{f_i\}_{i=1}^{\infty}$ and a real sequence of constant coefficients $\{a_i \neq 0\}_{i=1}^{\infty}$ decompose Kolmogorov Complexity, if the following inequality holds for all inputs x and at least one fixed integer n:

$(n-1) K(x) \leq a_1 K(f_1(x)) + a_2 K(f_2(x)) + \ldots + a_n K(f_n(x)) + O(\log |x|)$ , $\forall x, \exists n \in \mathbb{N}$

$O(log(n))$ or $O(1)$   $|x|$

$\{a_i \neq 0\}_{i=1}^{\infty}$

$\{f_i\}_{i=1}^{\infty}$

$$(n-1)K(x) \leq a_1 K(f_1(x)) + a_2 K(f_2(x)) + \ldots + a_n K(f_n(x)) + O(\log|x|) \ , \ \forall x, \exists n \in \mathbb{N}$$

**Remark 1.1**: *The right hand side might require addition of O(log(n)) or O(1). And also n is not |x| i.e. the length of input x. n is the number of computable functions.*

**Lemma 1.1**: Given a sequence of bounded computable functions $\{f_i\}_{i=1}^{\infty}$ and a bounded (from above and bounded from below) real sequence of constant coefficients $\{a_i\}_{i=1}^{\infty}$ decomposing Kolmogorov Complexity, then the following inequalities hold:

$$(n-1)K(x) \leq a_1 K(f_1(x)) + a_2 K(f_2(x)) + \ldots + a_n K(f_n(x)) + O(\log|x|) \leq$$
$$n(K(x) + M)\sup(\{a_i\}_{i=1}^{\infty}) + O(\log|x|), \ \exists n \in \mathbb{N} \quad (1.1)$$

where M is the program upper bound for $\{f_i\}_{i=1}^{\infty}$.

**Proof**: Direct consequence of Theorem A.4 and definition 1.1 :

$$K(f_i(x)) \leq K(x) + K(f_i) \leq K(x) + M \quad (1.2)$$

Therefore

$$a_i K(f_i(x)) \leq (K(x) + M)\sup(\{a_i\}_{i=1}^{\infty}) \quad (1.3)$$

Adding up all the n inequalities:

$$\sum_{i=1}^{n} a_i K(f_i(x)) \leq n(K(x) + M)\sup(\{a_i\}_{i=1}^{\infty}) \quad (1.4)$$

By definition 1.1 again

$$(n-1)K(x) \leq \sum_{i=1}^{n} a_i K(f_i(x)) + O(\log|x|) \leq n(K(x) + M)\sup(\{a_i\}_{i=1}^{\infty}) + O(\log|x|) \quad (1.5)$$

□

As direct consequence of Lemma 1.1:

**Theorem 1.1**: Assume there is decomposition by $\{f_i\}_{i=1}^{\infty}$ and $\{a_i\}_{i=1}^{\infty}$ then

**a)** As $K(x) \longrightarrow \infty$

$$(n-1) \leq \frac{\sum_{i=1}^{n} a_i K(f_i(x))}{K(x)} \leq n\sup(\{a_i\}_{i=1}^{\infty}) \quad (1.6)$$

**b)** As $n \longrightarrow \infty$

$$K(x) \leq \frac{\sum_{i=1}^{n} a_i K(f_i(x))}{n} \leq (K(x) + M)\sup(\{a_i\}_{i=1}^{\infty})$$

$K(x) \longrightarrow \infty \qquad n \longrightarrow \infty \qquad \sup(\{a_i\}_{i=1}^{\infty})$

$$(n-1) \leq \frac{\sum_{i=1}^n a_i K(f_i(x))}{K(x)} \leq n \sup(\{a_i\}_{i=1}^\infty)$$

$$K(x) \leq \frac{\sum_{i=1}^n a_i K(f_i(x))}{n} \leq (K(x) + M) \sup(\{a_i\}_{i=1}^\infty) \quad (1.7)$$

**c)** As $K(x) \longrightarrow \infty$ and $n \longrightarrow \infty$ and assuming $\sup(\{a_i\}_{i=1}^\infty) = 1$ then

$$\frac{\sum_{i=1}^n a_i K(f_i(x))}{n K(x)} \longrightarrow 1 \quad (1.8)$$

All inequalities above are exact i.e. there are no O notation added to the right handside.

**Remark 1.1**: *The author hopes to find formulations as below, however it is not clear at this time if the limit in (1.9) exists due to the addition of the O-notation to the inequalities.*

*As direct consequence of the definition 1.2 and Theorem 1.1 case c:*

**(putative) Corollary 1.1**: As $n \longrightarrow \infty$ and decomposition 1.2 holds and $a_i \neq 0$

$$\left| K(x) - \sum_{i=1}^n \frac{a_i}{(n-1)} K(f_i(x)) \right| \longrightarrow 0, \quad \forall\, x, n \in \mathbb{N} \quad (1.9)$$
$$\left| K(x) - \sum_{i=1}^n \frac{a_i}{n} K(f_i(x)) \right| \longrightarrow 0, \quad \forall\, x, n \in \mathbb{N} \quad (1.10)$$

*This gives a good impression of the concept of decomposition i.e. $K(x)$ is decomposed by a large sum of smaller expressions in Kolmogorov Complexity.*

## 2. Geometric Linkage: Canonical Projections

One of the simplest non-trivial examples of decompositions is projection:

1. Assume input x is actually a tuple or $x = (w_1, w_2, \ldots, w_n)$
2. $f_i(w_1, w_2, \ldots, w_n) = (w_1, \ldots w_{i-1}, w_{i+1} \ldots, w_n)$ i.e. a computable projection that removes the ith member of the tuple. Let's call these particular computable projections $\pi_i$.
3. $a_i = 1$

Let's see if these assumptions form a decomposition:

For $n = 2$ it is known (Theorem A.3):

$$K(w_1, w_2) \leq K(w_1) + K(w_2)$$

$\pi_1(w_1, w_2) = w_2$
$\pi_2(w_1, w_2) = w_1$

$$n - 1 = 2 - 1 = 1 \quad K(w_1, w_2) \quad a_i = 1,\ i = 1, 2$$

$n = 3$

$n = 2$

$K(w_1, w_2) \leq K(w_1) + K(w_2)$

$\pi_1(w_1, w_2) = w_2$

Here the coefficient $n - 1 = 2 - 1 = 1$ for $K(w_1, w_2)$ and $a_i = 1$, $i = 1, 2$.

For $n = 3$ it is known [1, 3]:

$2 K(w_1, w_2, w_3) \leq K(w_1, w_2) + K(w_2, w_3) + K(w_1, w_3)$ (2.1)

$\pi_1(w_1, w_2, w_3) = (w_2, w_3)$
$\pi_2(w_1, w_2, w_3) = (w_1, w_3)$
$\pi_3(w_1, w_2, w_3) = (w_1, w_2)$

Here the coefficient $n - 1 = 3 - 1 = 2$ for $K(w_1, w_2, w_3)$ and $a_i = 1$, $i = 1, 2, 3$.

The author has not seen the generalization of the above but attempts to prove one here:

Convention: Let n = 4, m = 3, assuming ',' and '(' and ')' are coded in binary, then

$x = ((w_{1,1}, w_{1,2}, w_{1,3}, w_{4,4}), (w_{2,1}, w_{2,2}, w_{2,3}, w_{2,4}), (w_{3,1}, w_{3,2}, w_{3,3}, w_{3,4}))$

Apply the projection $\pi_2$

$\pi_2(x) = ((w_{1,1}, w_{1,3}, w_{4,4}), (w_{2,1}, w_{2,3}, w_{2,4}), (w_{3,1}, w_{3,3}, w_{3,4}))$

**Theorem 2.1**: Assume $x = (w_1, w_2, \ldots, w_m)$ and let $w_i = (w_{i,1}, w_{i,2}, \ldots, w_{i,l})$ where $n$ and $l$ are any fixed integers $l \geq n$, and $K(w_{i,j})$ is bounded from above

$(n - 1) K(x) \leq K(\pi_1(x)) + K(\pi_2(x)) + \ldots + K(\pi_n(x)) + O(\log(m))$  $\forall x, \forall n \forall m \in \mathbb{N}$  (2.2)

where $\pi_i(x) = (\pi_i(w_1), \pi_i(w_2), \ldots, \pi_i(w_m))$, $i = 1, 2, \ldots, n$.

**Remark 2.1**: *Proof of Theorem 2.1 needs to be more thoroughly reviewed.*

**Proof by Induction**:

Assume $O(1)$ added to the right hand side of all inequalities.

Assume x is arbitrary as specified in the theorem and assume n is the induction parameter.

As shown above case n = 2, 3 are already proven. Case n = 1 is trivial.

**Case n**:

Assume the following inequality holds for n

$(n - 1) K(x) \leq K(\pi_1(x)) + K(\pi_2(x)) + \ldots + K(\pi_n(x)) + O(\log(m))$  (2.3)

$w_i C_i$ $\qquad$ $C_i$ $\qquad$ $w_i$

$(n-1) K(x) \leq K(\pi_1(x)) + K(\pi_2(x)) + \ldots + K(\pi_n(x)) + O(\log(m))$

**Case n + 1**:

Add a column C to each word in tuple x, so n is increased to n + 1 for the purpose of induction, and call the new input called 'xC'; by $w_i C_i$ we mean adding word $C_i$ to the end of a tuple $w_i$ :

$xC = (w_1 C_1, w_2 C_2, \ldots, w_m C_m)$  (2.4)

Let's prove the 'case n + 1' i.e. the following based upon 'case n' (2.3):

$n K(xC) \leq K(\pi_1(xC)) + K(\pi_2(xC)) + \ldots + K(\pi_n(xC)) + K(\pi_{n+1}(xC)) + O(\log(m))$  (2.5)

By Theorem A.4, and assuming a helper computable function $h_C(x) = xC$ :

$K(xC) \leq K(x) + K(h_C)$  (2.6)

Since $h_C$ adds a column of words, assuming n fixed, $K(h_C)$ has order of magnitude of $O(\log(m))$, because a loop of m rows needs to store loop-counter with value of m, which has length log(m), therefore:

$K(h_C) \leq K(h) + K(C) \leq O(\log(m)) + O(1)$  (2.7)

$K(C)$ is bounded from above therefore only contribute $O(1)$. And of course getting rid of the redundant O(1) notation:

$K(h_C) \leq O(\log(m))$  (2.8)

Therefore inequality (2.7) can be re-expressed

$K(xC) \leq K(x) + O(\log(m))$  (2.9)

Now use 'Case n' (2.3) instead with x = xC for some C:

$(n-1) K(xC) \leq K(\pi_1(xC)) + K(\pi_2(xC)) + \ldots + K(\pi_n(xC)) + O(\log(m))$  (2.10)

Now add the inequality (2.9) side by side to (2.10) and repeat $O(\log(m))$ only once:

$(n-1) K(xC) + K(xC) \leq K(\pi_1(xC)) + K(\pi_2(xC)) + \ldots + K(\pi_n(xC)) + K(x) + O(\log(m))$  (2.11)

However note

$K(\pi_{n+1}(xC)) = K(x)$  (2.12)

Finally

$n K(xC) \leq K(\pi_1(xC)) + K(\pi_2(xC)) + \ldots + K(\pi_n(xC)) + K(\pi_{n+1}(xC)) + O(\log(m))$  (2.13)

$$K(\pi_{n+1}(xC)) = K(x)$$

Consequently Case n + 1 is also correct, therefore by process of mathematical induction we proved for all n inequality (2.2) holds.

□

**Remark 2.2**: *Any other ordering of the elements of the tuples e.g. placing all index i into one string/word [3] does not change the outcome of the Theorem 2.1 since any such 'sorting' is computable and by Theorem A.4 at most O(1) is added to the inequality.*

# 3. Study: Complexity Analysis of Light Cone

The author was asked to produce practical studies of actual concepts in sciences in order to shed light on the applicable usages of Kolmogorov Complexity and the proposed decomposition. For that matter a simple study of the popular Light Cone from General Relativity has been used, however what follows is free of all smooth operators and their differential equations, hence no differential geometry.

This is a Light Cone [2]:

The cone itself is the locus of all points where the motion has speed of light.

The volume inside the cone is the locus of all points where the motion has speed less than that of light.

The outer surrounding volume is the locus of all points where the motion has speed higher than the speed of light.

FIG 3.1

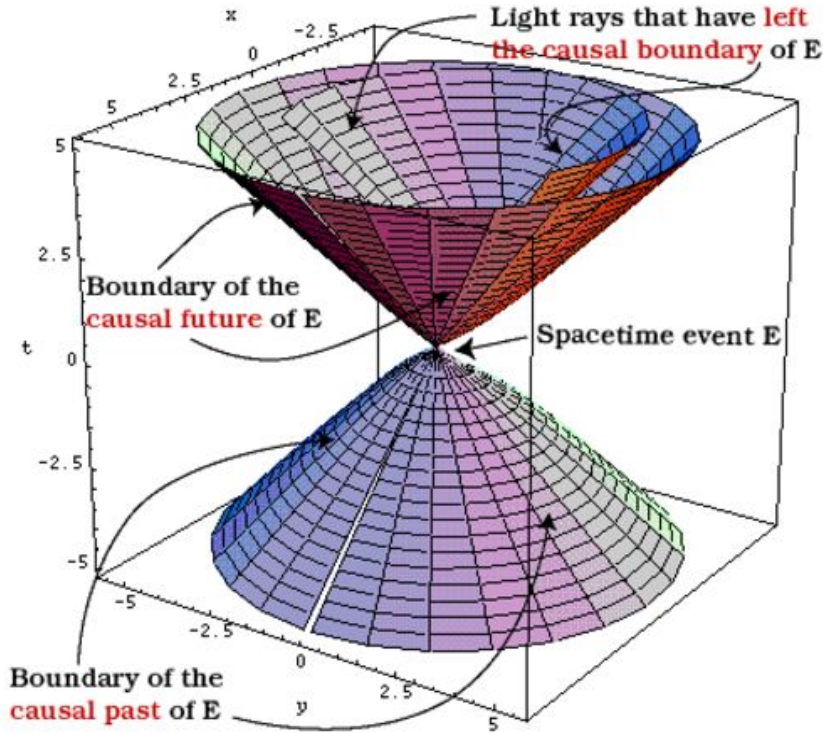

Normally all the above is captured algebraically as following:

$$\begin{cases} x^2 + y^2 + z^2 = c^2 t^2 & \text{on the cone} \\ x^2 + y^2 + z^2 < c^2 t^2 & \text{inside the cone} \\ x^2 + y^2 + z^2 > c^2 t^2 & \text{outside the cone} \end{cases} \quad (3.1)$$

Without loss of generality, per suggestion of Richard Feynman, it is more understandable to consider c = 1, and let's confine the space time to 1 second of past and 1 second of future i.e. the hypercube:

$[-1, 1] \times [-1, 1] \times [-1, 1] \times [-1, 1]$   (3.2)

Each point in this hypercube is a tuple (x, y, z, t) subject to the system of equations (3.1) assuming c = 1.

Space or Space Time is a collection of autonomous robotic probes who report on the local parameters of the space e.g. x or y.

Within the said hypercube we make a random 'indexed' grid of thousands of such autonomous robotic probes and they report back their space time coordinates, in an orderly fashion:

**The collective motion of these autonomous robotic probes is constrained: No matter how they move their coordinates' collective Kolmogorov Complexity is constant (preserved).**

$xyzt = \{(x_i, y_i, z_i, t_i)\}_{i=1}^{n}$

We then calculate the Kolmogorov Complexity of the Space Time as following:

Make sequences of each parameter into words/strings

$$\text{xyzt} = \{(x_i, y_i, z_i, t_i)\}_{i=1}^{n}$$

$$\begin{aligned}
\text{yzt} &= \pi_1(\text{xyzt}) = \{(y_i, z_i, t_i)\}_{i=1}^{n} \\
\text{xzt} &= \pi_2(\text{xyzt}) = \{(x_i, z_i, t_i)\}_{i=1}^{n} \\
\text{xyt} &= \pi_3(\text{xyzt}) = \{(x_i, y_i, t_i)\}_{i=1}^{n} \\
\text{xyz} &= \pi_4(\text{xyzt}) = \{(x_i, y_i, z_i)\}_{i=1}^{n}
\end{aligned} \quad (3.3)$$

$$\begin{aligned}
\text{zt} &= \pi_1(\pi_1(\text{xyzt})) = \{(z_i, t_i)\}_{i=1}^{n} \\
\text{yt} &= \pi_2(\pi_1(\text{xyzt})) = \{(y_i, t_i)\}_{i=1}^{n} \\
\text{yz} &= \pi_3(\pi_1(\text{xyzt})) = \{(y_i, z_i)\}_{i=1}^{n} \\
\text{xt} &= \pi_2(\pi_2(\text{xyzt})) = \{(x_i, t_i)\}_{i=1}^{n} \\
\text{xy} &= \pi_3(\pi_3(\text{xyzt})) = \{(x_i, y_i)\}_{i=1}^{n} \\
\text{xz} &= \pi_2(\pi_4(\text{xyzt})) = \{(x_i, z_i)\}_{i=1}^{n}
\end{aligned}$$

Then calculate

$K(\text{xyzt})$

**Remark 3.1**: $\text{xyzt} = \{(x_i, y_i, z_i, t_i)\}_{i=1}^{n}$ *as arbitrary points in space is serialized i.e. the hypothetical autonomous robots, no matter how they transmit their information, the receiver has to serialize their reception one way or another, in order to store an process.*

For example we randomly scattered 40,000+ such probes and obtained the Kolmogorov Complexity

$$K(\text{xyzt}) = 1,753,224 \quad (3.4)$$

FIG 3.2

$K(xyzt) = 1,753,224$

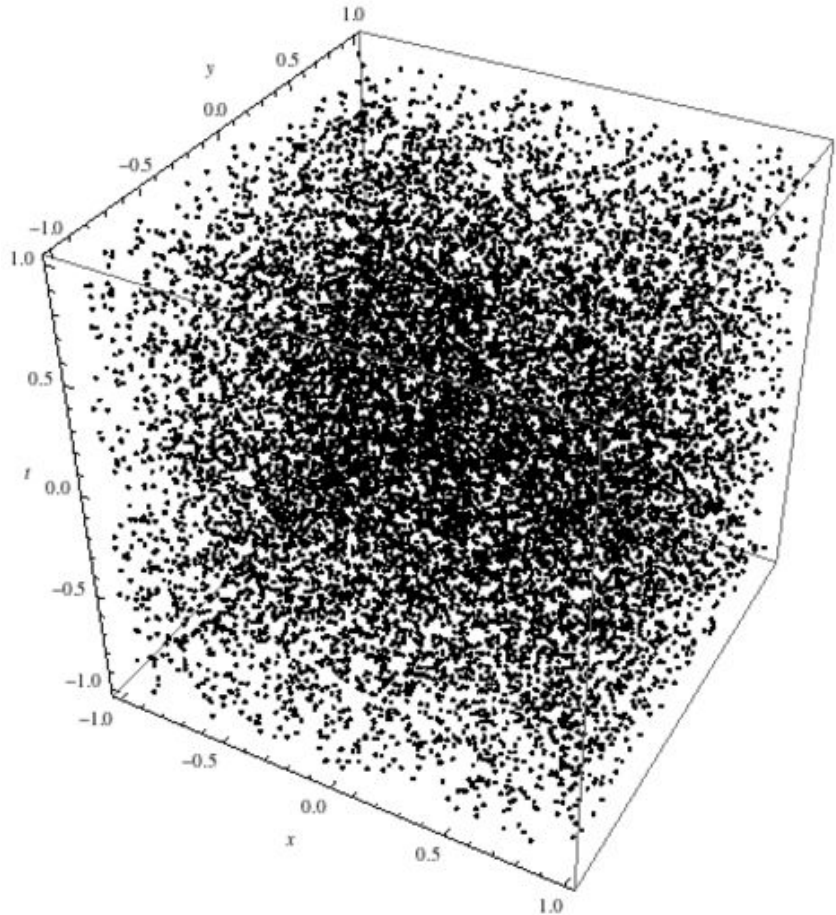

The we calculate the complexity of the projections

$K(xyz) = 1,324,160$
$K(yzt) = 1,323,952$      (3.5)
$K(xzt) = 1,324,208$
$K(xyt) = 1,324,248$

**Observation 1**: *Projections are much less in Kolmogorov Complexity, but still substantial and indicate randomness free of any discernible patterns.*

**Observation 2**: *Projections are indistinguishable since their Kolmogorov Complexity are very close in numbers.*

We project one more time and obtain 2D configurations and their complexities, for example:

$K(xt) = 890,488$      (3.6)

Again the Kolmogorov Complexity is dropped in amount but still substantial and thus indicate randomness, as we can see in the plot of the 2D projection into x-t plane:

FIG 3.3

$K(\text{xt}) = 890, 488$

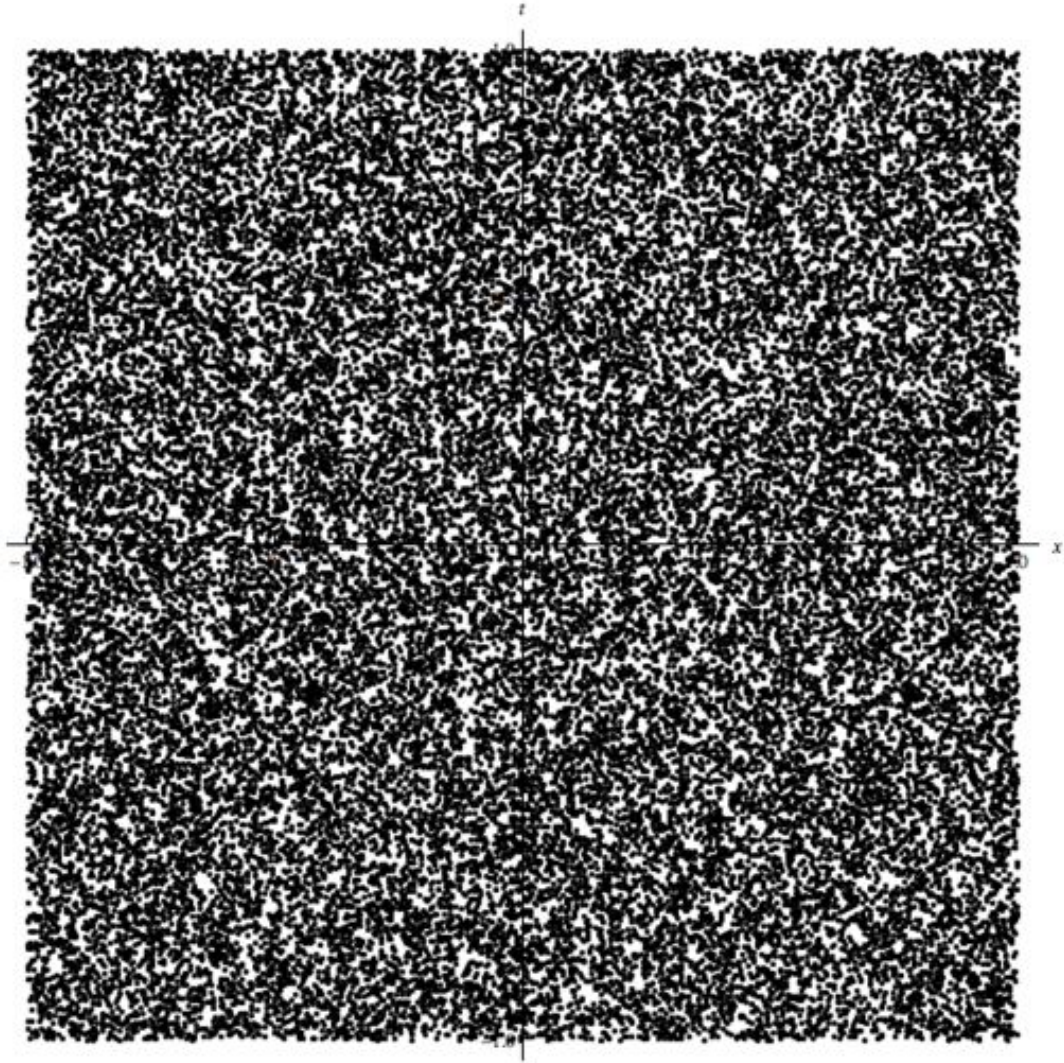

Note that by Theorem A.4

$$K(\pi_i(x)) \leq K(x) + K(\pi_i) \quad (3.7)$$

Generally the Kolmogorov Complexity of $\pi_i(x)$ drops by $K(\pi_i)$ amount which is small, but noticeable.

### Inside The Cone: Causality, Low Kolmogorov Complexity

Assuming c = 1 the interior of the of the cone is defined by the inequality:

$$x^2 + y^2 + z^2 < c^2 t^2 \implies x^2 + y^2 + z^2 < 1^2 t^2 \implies x^2 + y^2 + z^2 - t^2 < 0 \quad (3.8)$$

$K(\text{xyzt}) = 223, 344$

$$x^2 + y^2 + z^2 < c^2 t^2 \implies x^2 + y^2 + z^2 < 1^2 t^2 \implies x^2 + y^2 + z^2 - t^2 < 0$$

Preserving the order of original scattered points in the hypercube, apply the right hand side of the Eq (3.8) as a predicate to test for which points are inside the cone and then find the Kolmogorov Complexity in an orderly fashion i.e. original order preserved:

$$K(xyzt) = 223,344 \quad (3.9)$$

Compare this number to (3.4) the Kolmogorov Complexity for the entire hypercube 1,753,224.

Clearly substantial amount of reduction in Kolmogorov Complexity of the entire space compared to the Kolmogorov Complexity of the interior of the Light Cone!

**Observation 3**: *Therefore if we build a low-pass filter for Kolmogorov Complexity: The observable universe with lower than speed of light motion, passes through the low-pass filter, while the entire space of the hypercube will be occluded i.e. we see only the interior of the Light Cone.*

By looking at the projections of the interior of the Light Cone, e.g. a 2D plate x-y combine by measuring time i.e. the 3D space (x, y, t), we can observe further drop as the result of Eq (3.7):

$$K(xyt) = 168,240 \quad (3.10)$$

Clearly the low Kolmogorov Complexity indicates the appearance of the patterns i.e. the cone shaped volume!

FIG 3.4

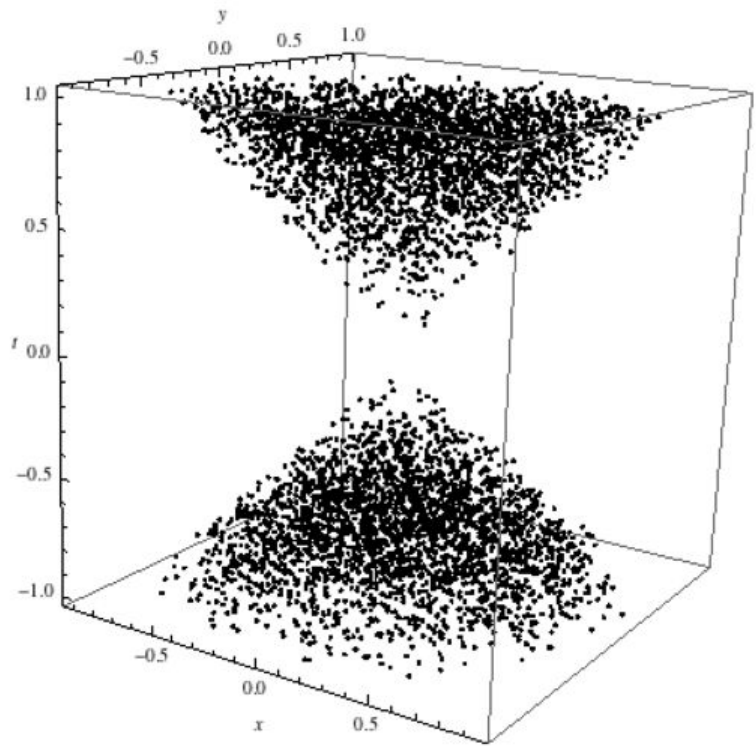

And apply the Projection even further results into more reduction in Kolmogorov Complexity:

$$K(xt) = 112,424$$

$$K(xyzt) = 1,753,224$$

$K(\text{xt}) = 112,424$ (3.11)

Compare this to the Eq (3.4) with the amount $K(\text{xyzt}) = 1,753,224$ for the Kolmogorov Complexity of the entire hypercube, we can see how sharp the drop in Kolmogorov Complexity is.

Clearly the low Kolmogorov Complexity indicates the appearance of patterns i.e. the conic section!

FIG 3.5

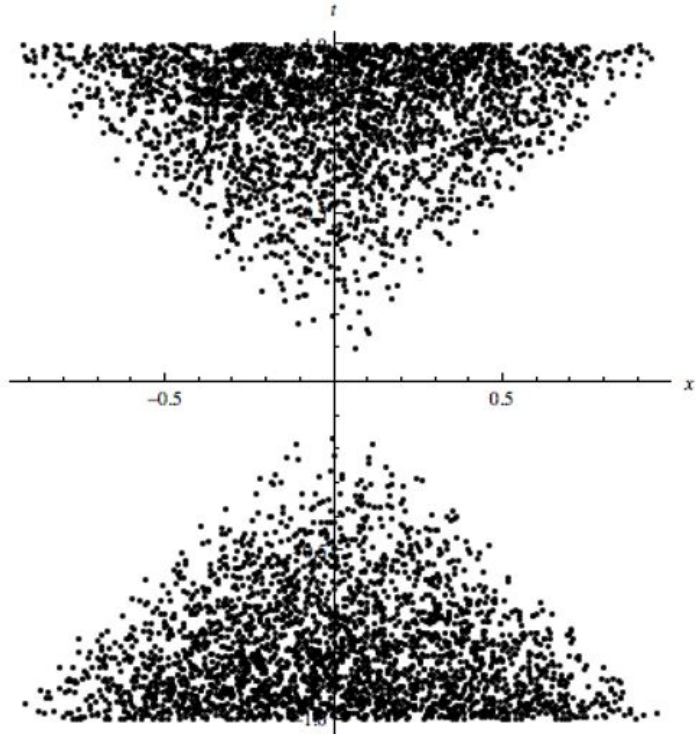

Clearly the low Kolmogorov Complexity indicates the appearance of the patterns i.e. the spherical volume!

$K(\text{xyz}) = 170,344$ (3.12)

**Observation 4**: *Projections are distinguishable albeit their Kolmogorov Complexity being quite close in numbers, due to the constrained imposed by the inequality of the Light Cone.*

FIG 3.6

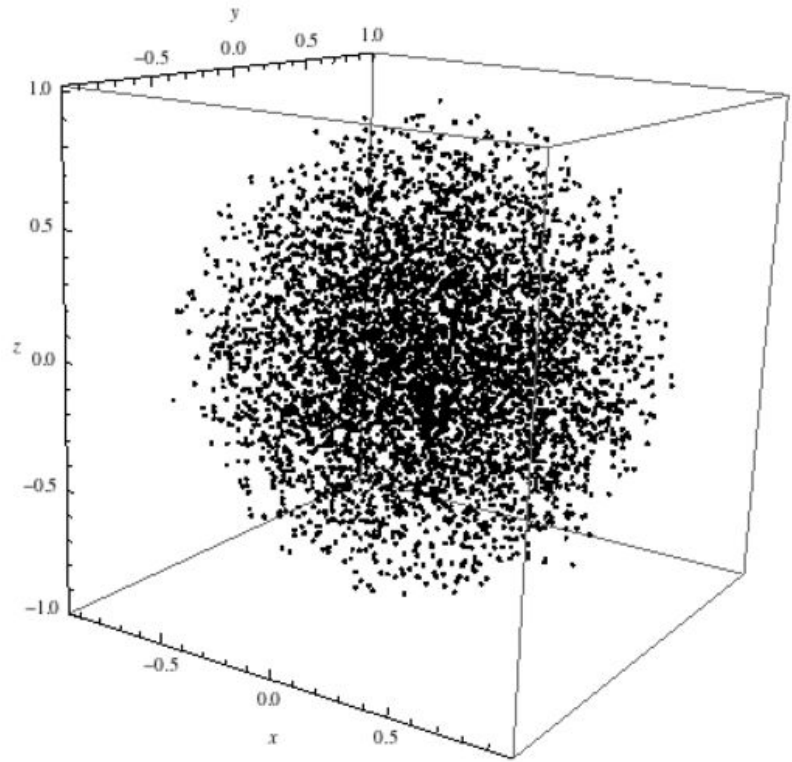

Clearly the low Kolmogorov Complexity indicates the appearance of the patterns i.e. the disk area!

$K(xy) = 114, 592$     (3.13)

FIG 3.7

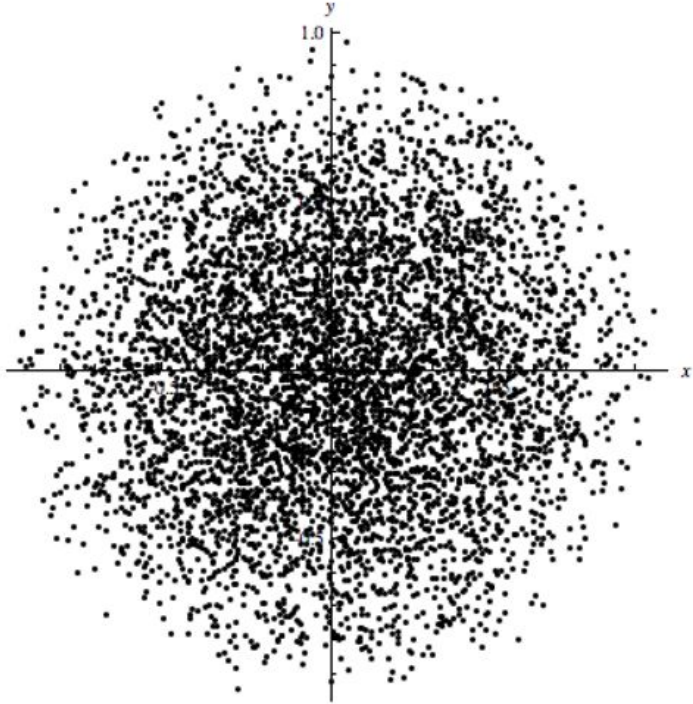

### Outside The Cone: Non-Causality, High Kolmogorov Complexity

Equation for screening the points in the hypercube, assuming c = 1:

$$x^2 + y^2 + z^2 - t^2 > 0 \quad (3.14)$$

These points indicate motion faster than light.

The Kolmogorov Complexity of the outside of the Light Cone:

$$K(xyzt) = 1,528,344 \quad (3.15)$$

Amazingly comparing this number to Eq (3.4) 1, 753, 224 has reduced by relatively small amount, as opposed to the corresponding number within the Light Cone Eq (3.9) $K(xyzt) = 223,344$.

And projecting the outside volume into x-y-t space, the reduced Kolmogorov Complexity generates a hazy pattern:

$$K(xyt) = 1,154,632 \quad (3.16)$$

FIG 3.8

$K(xyt) = 1, 154, 632$ 

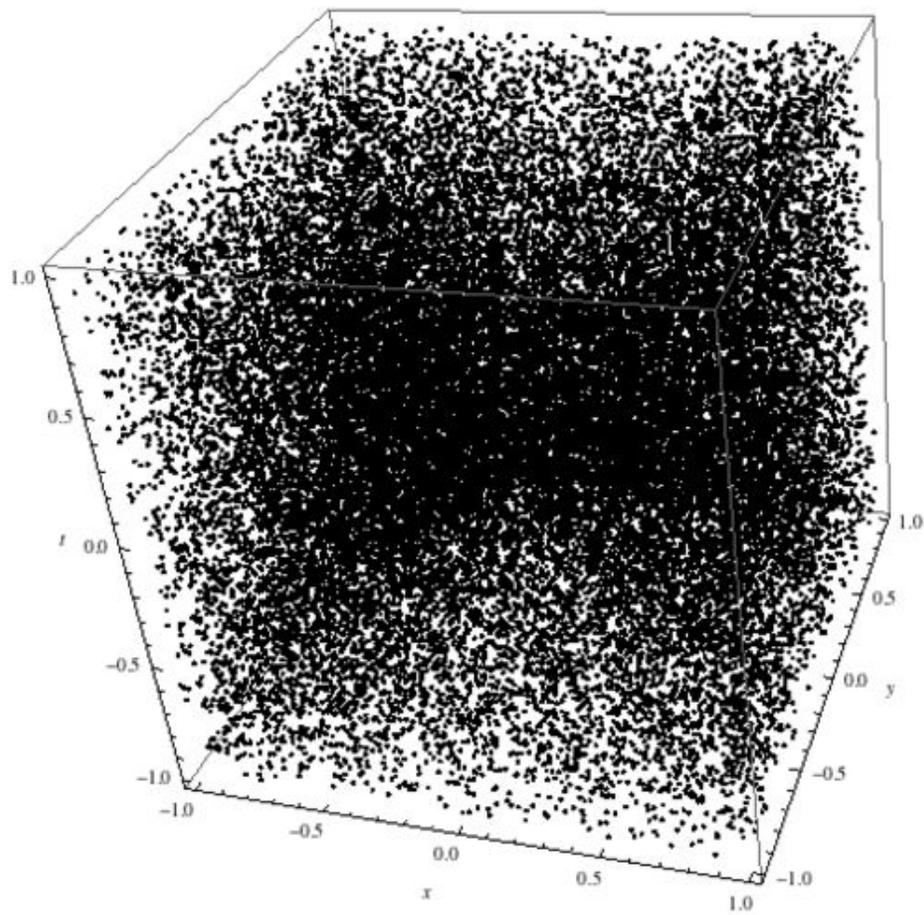

Reduced Kolmogorov Complexity shows slight lowering of the distribution of the points in a hazy conic section area.

$K(xt) = 776, 848$     (3.17)

FIG 3.9

$K(\text{xt}) = 776, 848$

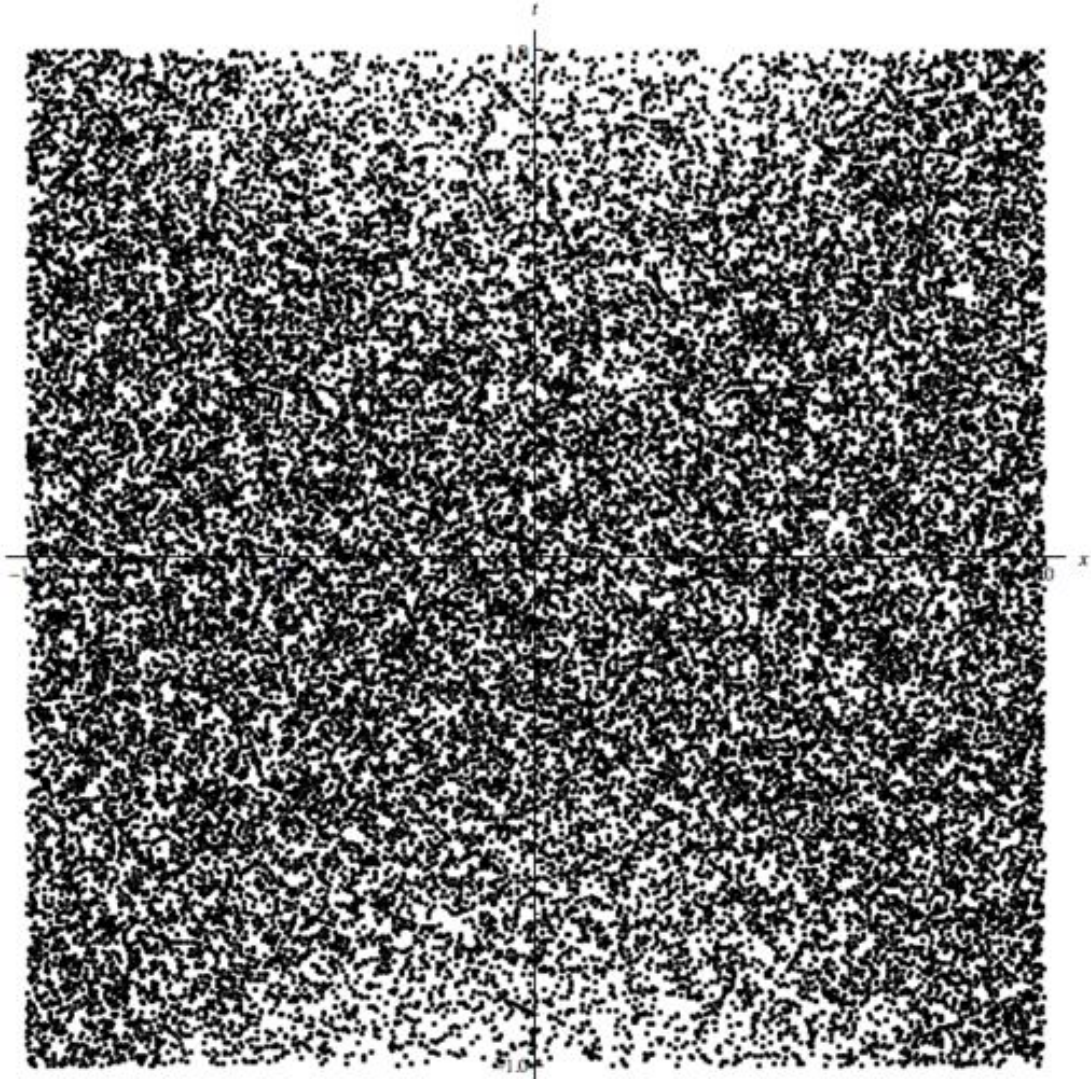

$K(\text{xyz}) = 1, 153, 160$ \qquad (3.18)

FIG 3.10

$K(xyz) = 1, 153, 160$

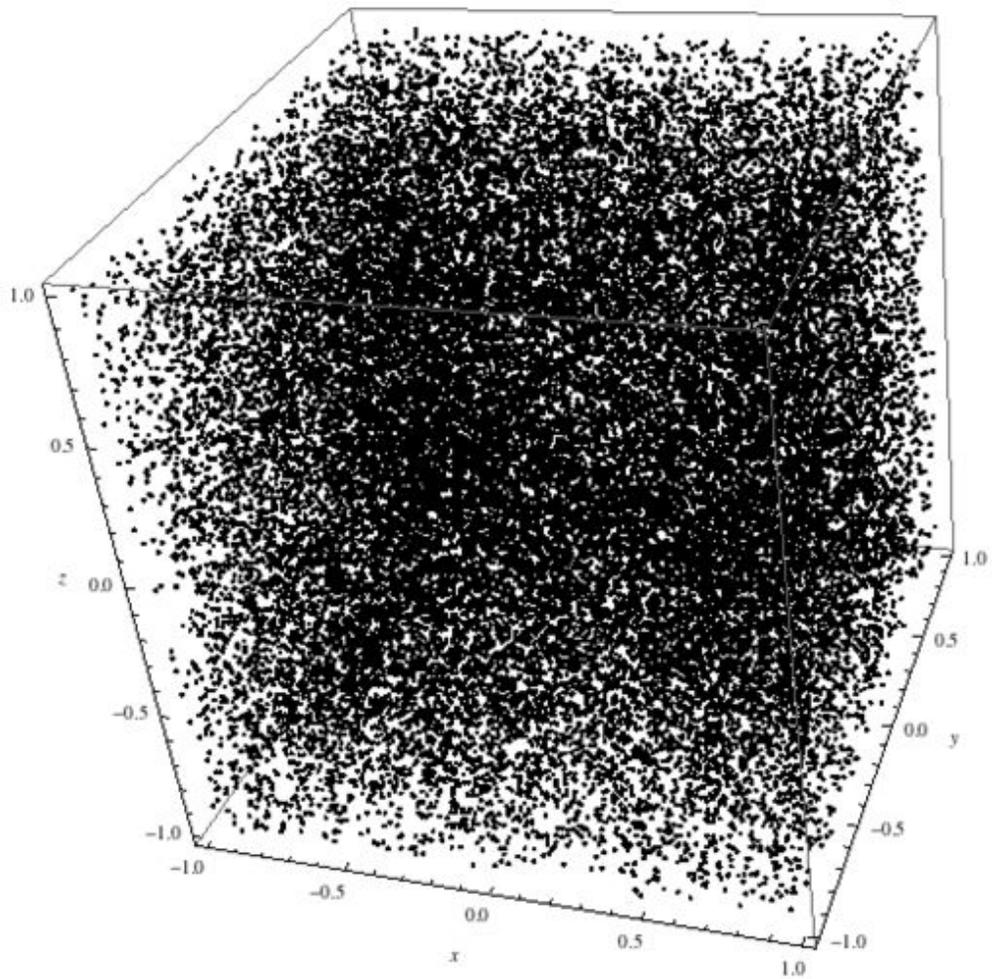

Reduced Kolmogorov Complexity shows slight lowering of the distribution of the points in a hazy disk area.

$K(xy) = 775, 248$ (3.19)

FIG 3.11

$K(xy) = 775, 248$

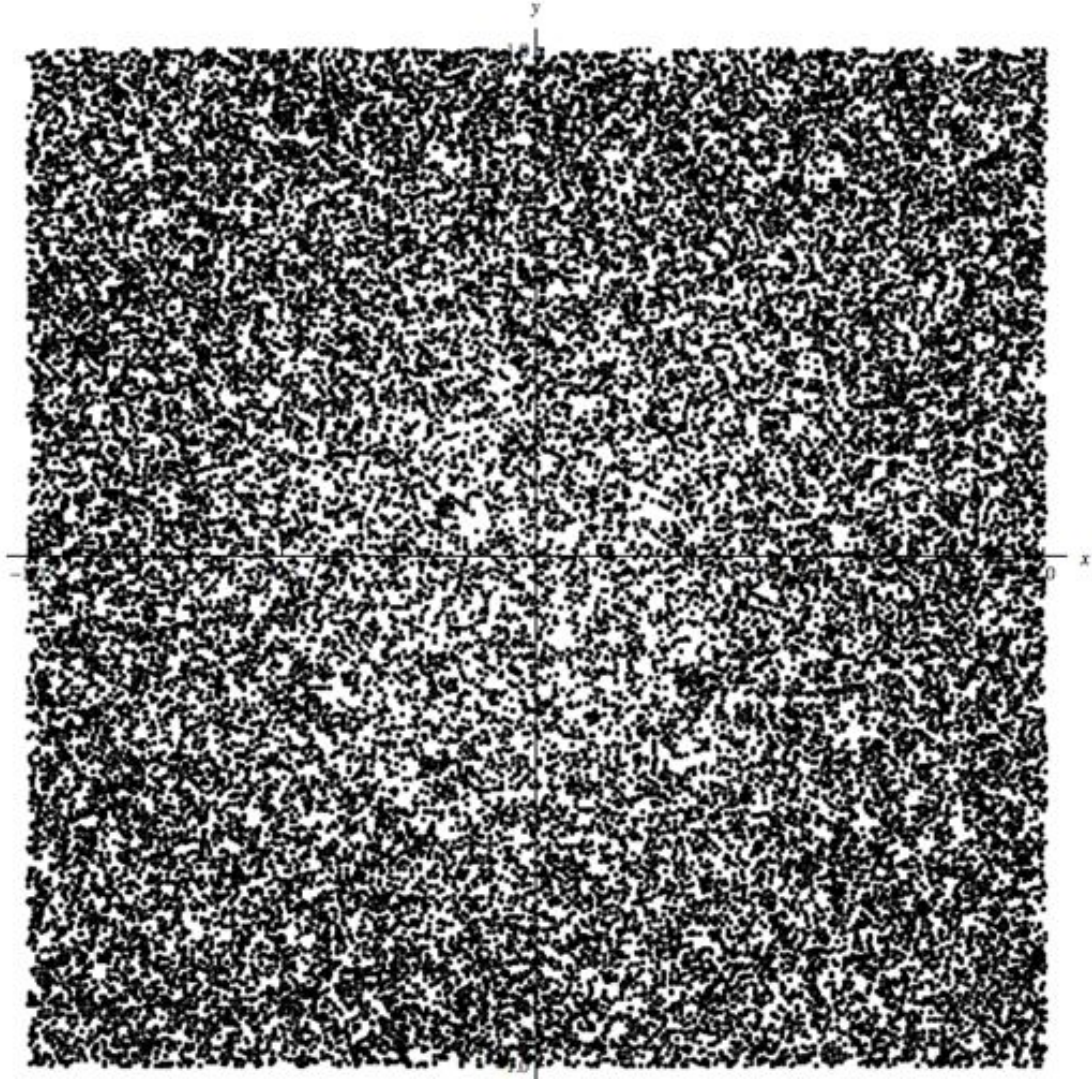

**Observation 5**: *Therefore if we build a High-pass filter for Kolmogorov Complexity: The un-observable universe with faster than speed of light motion, passes through the High-pass filter, like a foam ocean of Brownian Motions i.e. the outside of the Light Cone, the non-Causal universe. Hazy foggy patterns are discernible at best.*

# Appendix A

## Kolmogorov Complexity

$$C_{\mathcal{U}}(x)$$

$$C_{\mathcal{U}}(x) = \min_{p:\mathcal{U}(p) = x} l(p)$$

We assume all strings and programs are binary coded.

**Definition A.1**: The Kolmogorov Complexity $C_\mathcal{U}(x)$ of a string x with respect to a universal computer (Turing Machine) $\mathcal{U}$ is defined as

$$C_\mathcal{U}(x) = \min_{p:\mathcal{U}(p)=x} l(p)$$

the minimum length program p in $\mathcal{U}$ which outputs x.

**Theorem A.1 (Universality of the Kolmogorov Complexity)**: *If $\mathcal{U}$ is a universal computer, then for any other computer $\mathcal{A}$ and all strings x,*

$$C_\mathcal{U}(x) \leq C_\mathcal{A}(x) + c_\mathcal{A}$$

where the constant $c_\mathcal{A}$ does not depend on x.

**Corollary A.1**: $\lim_{l(x) \to \infty} \frac{C_\mathcal{U}(x) - C_\mathcal{A}(x)}{l(x)} = 0$ *for any two universal computers.*

**Remark A.1**: *Therefore we drop the universal computer subscript and simply write $C(x)$.*

**Definition A.2**: Self-delimiting string (or program) is a string or program which has its own length encoded as a part of itself i.e. a Turing machine reading Self-delimiting string knows exactly when to stop reading.

**Definition A.3**: The Conditional or Prefix Kolmogorov Complexity of self-delimiting string x given string y is

$$K(x \mid y) = \min_{p:\mathcal{U}(p,y)=x} l(p)$$

The length of the shortest program that can compute both x and y and a way to tell them apart is

$$K(x, y) = \min_{p:\mathcal{U}(p)=x,y} l(p)$$

**Remark A.2**: *x, y can be thought of as concatenation of the strings with additional separation information.*

**Theorem A.2**: $K(x) \leq l(x) + 2 \log l(x) + O(1), \quad K(x \mid l(x)) \leq l(x) + O(1)$.

**Theorem A.3**: $K(x, y) \leq K(x) + K(y)$.

**Theorem A.4**: $K(f(x)) \leq K(x) + K(f)$ , $f$ a computble function

# Appendix B

```
(* Approximates the Kolmogorov Complexity function on arbitrary input *)
kolComplexity[x0_] := Module[{x = x0, xdelim, compxdelim},
```

# Approximate Kolmogorov Complexity in *Mathematica*

```
(* Approximates the Kolmogorov Complexity function on arbitrary input *)
kolComplexity[x0_] := Module[{x = x0, xdelim, compxdelim},

    (* delimit the input x *)
    xdelim = {x, ByteCount[x]};

    (* Compress xdelim and count the bytes of the reuslting compression *)
    compxdelim = ByteCount[Compress[xdelim]];

    (* Return the tuple {K(delimx),|delimx|}*)
    {compxdelim, ByteCount[xdelim]}
]
```

Let's test the above Module; make a very long string x such that is all 1s, 2^20 = 1,048,576:

```
x = Table[1, {i, 1, 2^20}];
kolComplexity[x]
```

{5544, 4 194 552}

First number 5,544 is its approximated Kolmogorov Complexity, the second number 4,194,552 is the length of original input x delimited. Length is measured in bytes.

Now introduce some complexity i.e. make x a very long alternating string of 1s and 0s:

```
x = Table[Boole[EvenQ[i]], {i, 1, 2^20}];
kolComplexity[x]
```
{8256, 4 194 552}

Kolmogorov Complexity is increased! from 5,544 to 8,256.

Now let's turn up the complexity by making x an almost random string of bits:

```
x = Table[RandomInteger[], {i, 1, 2^20}];
kolComplexity[x]
```
{351 496, 4 194 552}

Kolmogorov Complexity shoots up from 8,256 to 351,496 about 2 orders of magnitude!

Let's verify the theorems.

```
n = 100;
m = 3000;
(* Make a random sequence of m points in a hypercube of dimension n *)
pts = randomGrid[hyperCube[n], m];

k = kolComplexity[pts]

(n - 1) * k[[1]]
(* Find all the projections *)
decomposition = Table[Drop[pts, None, {i, i}], {i, 1, n}];

(* Sum the Kolmogorov Complexity of the decomposed projections *)
decsum =
 Total[Table[ kolComplexity[decomposition[[i]]][[1]], {i, 1, Length[decomposition]}] ]

(* Calculate   (n-1) ≤ (∑ᵢ₌₁ⁿ aᵢK(fᵢ(x)))/K(x)   (1.6)   *)
N[decsum / k[[1]]]
```

{3 150 864, 9 744 120}

311 935 536

311 936 832

99.0004

Eq (1.6) in Theorem 1.1 is verified where $n - 1 \approx 99.0004$ where $n = 100$. Also Eq (2.2) of Theorem 2.1 is verified by:

$(n - 1) K(x) \leq K(\pi_1(x)) + K(\pi_2(x)) + \ldots + K(\pi_n(x))$

$(n - 1) K(x) \approx 311{,}935{,}636$

$K(\pi_1(x)) + K(\pi_2(x)) + \ldots + K(\pi_n(x)) \approx 311{,}936{,}832$

Same verification was applied to points on a parametric curve within the hypercube i.e. to have a reduced Kolmogorov Complexity, about an order of magnitude reduction, example:

$K(\pi_1(x)) + K(\pi_2(x)) + \ldots + K(\pi_n(x)) \approx 311,936,832$

```
n = 100;
m = 3000;
(*pts=randomGrid[hyperCube[n], m];*)
zeros = ConstantArray[0, n - 3];
pts =
  Flatten[Table[Table[Join[{a*Cos[t], (a^2)*Sin[t], a}, zeros], {t, 0, 2*Pi, 0.1}],
    {a, -1, 1, 0.01}], 1];

k = kolComplexity[pts]

(n - 1) * k[[1]]
decomposition = Table[Drop[pts, None, {i, i}], {i, 1, n}];
decsum =
 Total[Table[ kolComplexity[decomposition[[i]]][[1]], {i, 1, Length[decomposition]}] ]
N[decsum / k[[1]]]
```

{350 320, 41 129 544}

34 681 680

34 748 816

99.1916

$n - 1 \approx 99.1916$ where $n = 100$

$(n - 1) K(x) \leq K(\pi_1(x)) + K(\pi_2(x)) + \ldots + K(\pi_n(x))$

$(n - 1) K(x) \approx 34,681,680$

$K(\pi_1(x)) + K(\pi_2(x)) + \ldots + K(\pi_n(x)) \approx 34,748,816$

# References

[1] M. Li and P. M. B. Vitanyi, Kolmogorov complexity and its applications, in: J. van Leeuwen (ed.), Handbook of Theoretical Computer Science, vol. A, Elsevier, Amsterdam, 1990, pp. 187–254.

[2] http://themaclellans.com/timetravel.html

[3] D. Hammer and A. Shen, A Strange Application of Kolmogorov Complexity, Theory Comput. Systems 31, 1–4 (1998)